\documentclass{article}
\usepackage{spconf,amsmath,graphicx, multirow, xcolor}
\usepackage{cite,url}

\title{Addressing the Confounds of Accompaniments in Singer Identification}

\name{Tsung-Han Hsieh$^{1,2}$, Kai-Hsiang Cheng$^{1}$, Zhe-Cheng Fan$^{1}$, Yu-Ching Yang$^{2}$, Yi-Hsuan Yang$^{1}$}
\address{$^{1}$~Research Center for IT Innovation, Academia Sinica, Taiwan  \\
$^{2}$~Data Science Team, KKBOX Inc., Taiwan \\
\small \tt \{bill317996,kevinco27,zcfan\}@citi.sinica.edu.tw, janetyang@kkbox.com, yang@citi.sinica.edu.tw}

\begin{document}
\maketitle
\begin{abstract}
Identifying singers is an important task with many applications. However, the task remains challenging due to many issues. One major issue is related to the confounding factors from the background instrumental music that is mixed with the vocals in music production. 
A singer identification model may learn to extract non-vocal related features from the instrumental part of the songs, if a singer only sings in certain musical contexts (e.g., genres). The model cannot therefore generalize well when the singer sings in unseen contexts.
In this paper, we attempt to address this issue.
Specifically, we employ \emph{open-unmix}, an open source tool with state-of-the-art performance in source separation, to separate the vocal and instrumental tracks of music. We then investigate two means to train a singer identification model: by learning from the separated vocal only, or from an augmented set of data where we ``shuffle-and-remix'' the separated vocal tracks and instrumental tracks of different songs to artificially make the singers sing in different contexts.  
We also incorporate melodic features learned from the vocal melody contour for better performance. 
Evaluation results on a benchmark dataset called the artist20 shows that this data augmentation method greatly improves the accuracy of singer identification.
\end{abstract}

\section{Introduction}
\label{sec:intro}
Singer identification (SID), a.k.a., artist identification, is a classic task in the field of music information retrieval (MIR). It aims at identifying the performing singers in given audio samples to facilitate management of music libraries. When properly trained, an SID model also learns the embedding of singing voices that can be used in downstream singing-related applications such as similarity search, playlist generation, or singing synthesis \cite{demetriouJKB18ismir,kaist19ismir,umbert15spm,liu19arxiv,humphrey19spm}. We refer readers to \cite{humphrey19spm} for a recent overview of research on singing voice analysis and processing, and the role of SID in related tasks.


\begin{figure}[t]
\begin{center}
\includegraphics[trim=5cm 0.5cm 0.5cm 2.5cm,clip,scale=0.40]{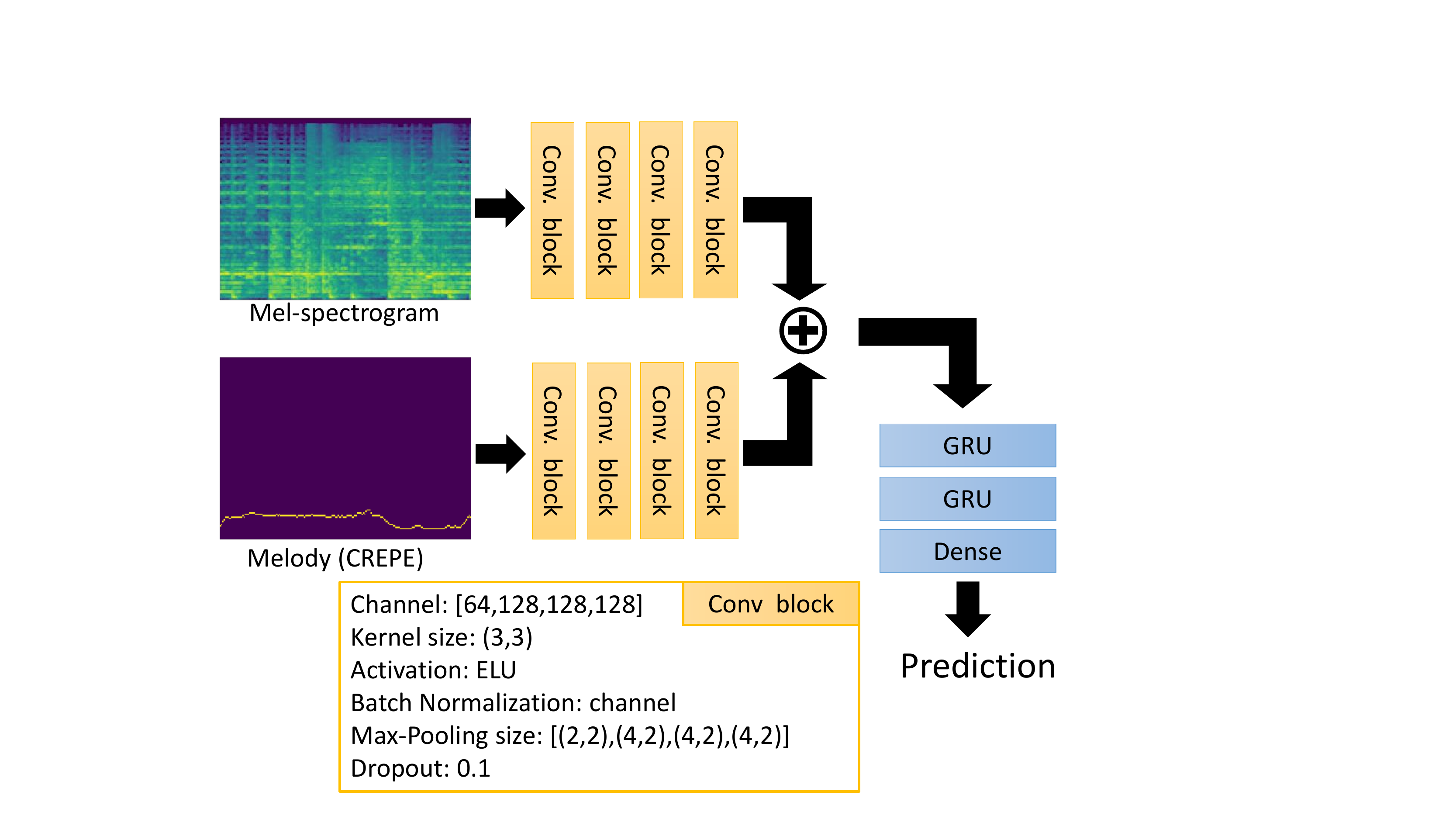}
\caption{The architecture of the proposed \emph{convolutional recurrent neural network with melody} (CRNNM) model for singer identification. The inputs are mel-spectrograms, and melody contours extracted by CREPE \cite{kim2018crepe}. The model cascades convolutional blocks, gate recurrent units (GRUs), and a dense layer. The ``+'' symbol stands for channel-wise concatenation.}
\label{fig:crnnm}
\end{center}
\end{figure}

Despite of its importance, SID is to date not yet a settled task \cite{humphrey19spm}.
There are at least two main challenges. First, as human beings share similar mechanism in producing sounds \cite{sundberg89book}, the difference in the singing voices of two singers may not be always obvious. 
This becomes more severe as the number of singers to be considered increases.
Second, due to the difficulty in acquiring solo recordings of singers, the training data for SID usually consists of audio recordings of singers singing over instrumental accompaniment tracks. 
The vocal track and instrumental track of a song are usually mixed in such an audio recording \cite{rafii18taslp}.
The presence of instrumental accompaniment not only makes it difficult for an SID model to extract only vocal-related features from the audio, but also introduces confounding factors \cite{sturm14mm} that hurt the model's generalizability. This is especially the case as singers usually have their preferred musical genres or styles. In trying to reproduce the most ground truth artist labels of a training dataset (e.g., while minimizing a classification error related loss function), an SID model may learn to capitalize non-vocal related features, which is not what the task is actually about.


We intend to address the second challenge in this paper. 
Intuitively, the challenge can be tackled by enhancing, or isolating out, the vocal part of a song, to minimize the effect of the instrumental part on the SID model. While singing voice enhancement or separation were difficult just a few years ago \cite{durrieu09icassp,sha13icassp}, 
state-of-the-art models now can perform the task with low distortion, interference and artifact  \cite{rafii18taslp,openunmix,liu19}, thanks to the advance in deep learning.
Using source separation (SS) to improve SID therefore becomes feasible.

While the idea of using SS to improve SID has been attempted before \cite{mesaros2007singer,sha13icassp,su13ismir,Sharma2019interspeech}, our work differs from the prior arts in two ways. First, except for the concurrent work \cite{Sharma2019interspeech},
the prior arts that we are aware of did not use deep learning-based SS models. In contrast, in our work both the SS model and the SID model  employ deep learning. Specifically, we use \emph{open-unmix} \cite{openunmix}, an open-source three-layer bidirectional deep recurrent neural network for SS.
Moreover, we build upon our SID model based on the implementation of a convolutional recurrent neural network made available by Nasrullah and Zhao \cite{nasrullah2019music}, which attains the highest song-level F1-score of 0.67 on the per-album split of the artist20 dataset \cite{artist20}, a standard dataset for SID.
As neural networks may find their own way extracting relevant features or patterns from the input, it remains to be studied whether the use of SS can improve the performance of a deep learning based SID model.
  
Second, unlike prior arts (including \cite{Sharma2019interspeech}), we investigate one additional way to employ SS to improve SID. Given the separated vocal tracks and instrumental tracks of the audio recordings in the training set, we perform the so-called ``data augmentation'' \cite{dataaug-intro,dataaug-is-work1,dataaug-is-work2, dataaug-JY}  by randomly \emph{shuffling} the separated tracks of different songs and then \emph{remixing} them.
For example, we remix the vocal part of a song from a singer with the instrumental part of another song from a different singer. 
In this way, we artificially make the singers sing over a variety of accompaniment tracks, and may therefore break the ``bonds'' between the vocal and accompaniment tracks, mitigating the confounds from the accompaniments.
We intend to empirically validate the effectiveness of such a data augmentation method, which can be said to be task-specific to SID.

As a secondary contribution, we explore adding to our SID model features extracted from the vocal melody contour, which is related to singing timbre \cite{panteli2017towards}. While the extraction of the vocal melody contour is done by using \emph{CREPE}~\cite{kim2018crepe},
an open-source tool with  state-of-the-art performance in melody extraction, we use a stack of convolutional layers and gated recurrent unit (GRU) layers~\cite{cho2014learning} to learn features not only from the mel-spectrogram but also the melody contour.\footnote{Features extracted from the melody contour have been shown useful in many other MIR tasks~\cite{Salamon2012icassp, Rocha2013ecml, panteli2017towards, Bittner2017AES}. However, we note that most existing work used hand-crafted features, rather than features learned by a neural network.}

Figure \ref{fig:crnnm} shows the architecture of our SID model, dubbed \emph{convolutional recurrent neural network with melody} (CRNNM). 
Code available at \url{https://github.com/bill317996/Singer-identification-in-artist20}.



\begin{figure}[t]
\begin{center}
\includegraphics[trim=0.01cm 0.01cm 0.01cm 0.01cm,clip,scale=0.33]{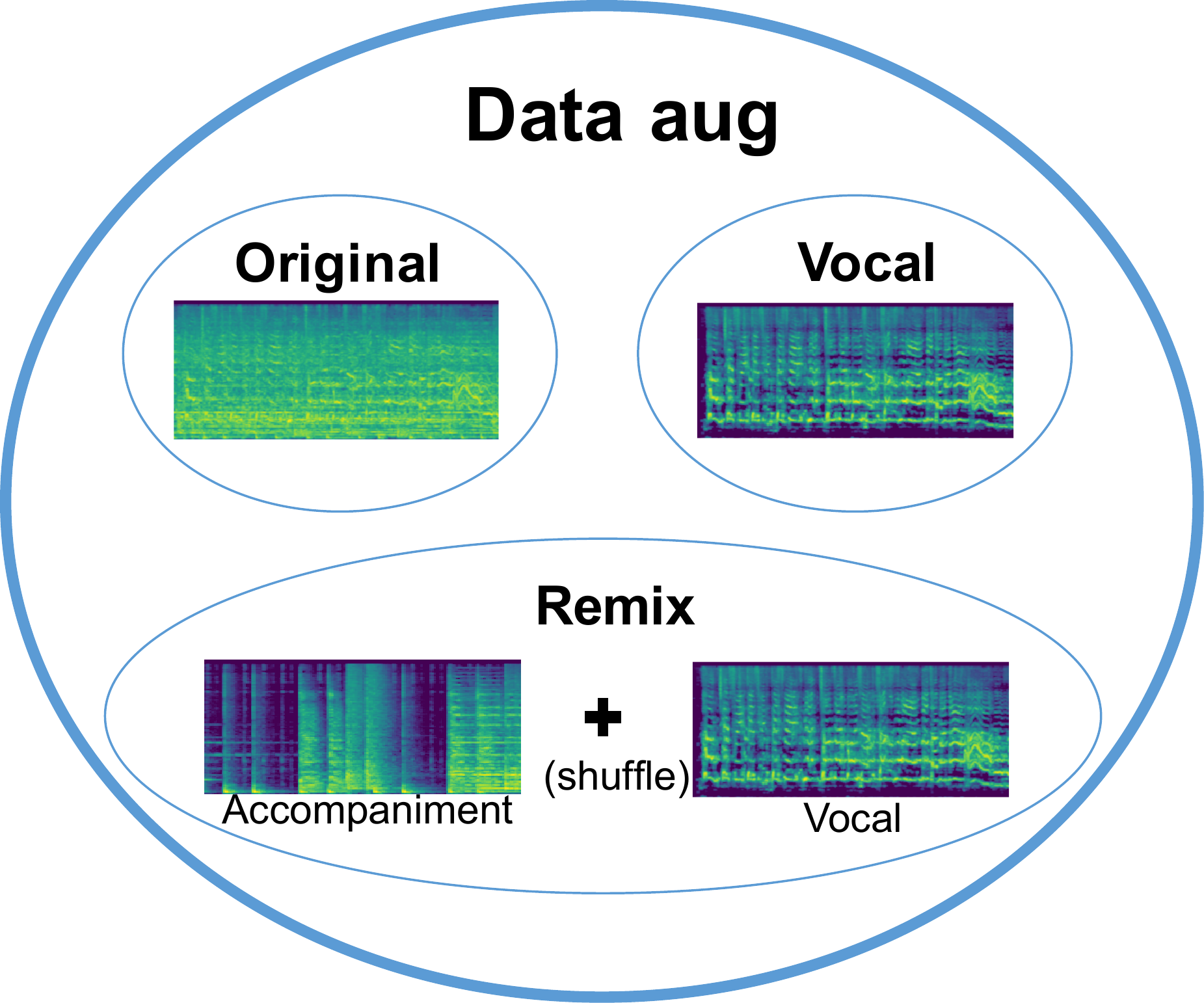}
\caption{A diagram of the ``shuffle-and-remix'' data augmentation method, which has been used before for SS \cite{dataaug-JY}.}
\label{fig:data_aug}
\end{center}
\end{figure}

\section{Methodology}
\label{sec:methodology}


\subsection{Singer Identification (SID) Models}

We consider as the baseline model the convolutional  recurrent neural network proposed in \cite{nasrullah2019music}, which represents the state-of-the-art for SID on the artist20 dataset. This model uses a stack of four convolutional layers, two GRU layers, and one dense (i.e., fully-connected) layer, as depicted in Fig. \ref{fig:crnnm}, but without the lower melody-related branch. We follow exactly the same design (i.e., encompassing number of filters, kernel sizes, activation functions, loss function, optimizer, learning rate, etc) of \cite{nasrullah2019music}.  We refer to this  model as `CRNN' below.

The proposed CRNNM model extends the CRNN model in two ways. First, in addition to the mel-spectrogram, we use CREPE \cite{kim2018crepe} to extract the melody contour from the mixture  audio recordings and establish an additional convolutional branch to learn melodic features for SID. For simplicity, we use the same design for the mel-spectrogram branch and the melody contour branch. 
Second, instead of using the mel-spectrogram of the mixture audio recordings, we employ open-unmix \cite{openunmix} to remove the instrumental part of the music, and use the proposed data augmentation technique to increase the size of the training data, as described below.


As CRNNM has more parameters than CRNN, in our experiment we also implement a variant of CRNN, denoted as CRNN$^\dagger$, that has similar number of parameters as CRNNM.

\subsection{Data Augmentation: Separate, Shuffle, and Remix}
\label{ssec:data_augmentation}


Data augmentation is to synthetically create training examples  to improve generalizability and to help capture invariances of data \cite{dataaug-intro}. This technique has been popular for some time among the machine learning community. It has also been shown beneficial for MIR tasks such as singing voice detection and source separation \cite{dataaug-is-work1, dataaug-is-work2, dataaug-JY} (but not yet for SID). 

As discussed in \cite{dataaug-is-work1}, data augmentation techniques for MIR can be classified into data-independent, audio-specific, and task-specific methods. Data-independent methods, like dropout, achieve augmentation from model perspective, and then can be data-agnostic. Audio-specific methods, like pitch shifting and time stretching, perform data transformation directly on audio data. 
Task-specific methods consider the task-specific prior knowledge into the training data. For example, it has been known that remixing sources from different songs improves the performance of SS models \cite{dataaug-JY}.


Our approach is motivated by \cite{dataaug-JY}. Our conjecture is that the same shuffle-and-remix technique can also be used for SID: when the vocal part of a song is mixed with the instrumental part of another song, its singer label should remain the same. This process is illustrated in Figure \ref{fig:data_aug}. Following this light, we create another three datasets, \emph{Vocals}, \emph{Remix}, and \emph{Data aug} to evaluate our model. 

\textbf{Origin}: The original audio recordings of artist20 \cite{artist20}. It contains six albums per artist for 20 artists, with in total 1,413 sound tracks. Vocal and acconamniments are mixed. 

\textbf{Vocal-only}: The vocal tracks separated by open-unmix \cite{openunmix}. In other words, all the accompaniments are removed.

\textbf{Remix}: The dataset is generated by randomly mixing the separated vocal and instrumental tracks of artist20. The size of this dataset is the same as Origin and Vocal-only.

\textbf{Data aug}: Combination of the  three sets above.

\subsection{Implementation Details}
\label{ssec:trainproc}

In the literature of SID, data splitting can be done in two  ways: song-split or album-split. The former splits a dataset by randomly assigning songs to the three subsets, whereas the latter makes sure that songs from the same album are either in the training, validation, or the test split.
It has been known \cite{nasrullah2019music} that song-split may leak production details associated with an album over the training and testing subsets, giving an SID model additional clues for classification.  Accordingly, the accuracy for song-split may be overly optimistic and tends to be higher than that of album-split.
We therefore focus on and only consider \textbf{album-split} in our work.

Under the album-split, we consider and compare the result of models trained using the four types of data listed by the end of Section \ref{ssec:data_augmentation}. The same test set (i.e., the Origin type) is used.

Following \cite{nasrullah2019music}, we cut the songs into 5-sec segments for training a 20-class classification model.
The final prediction result for a song is made by majority voting from the per-segment results.
For evaluation, we consider both ``per 5-sec segment''  and  ``per song'' F1 score; both the higher the better.

For CRNNM, we quantize the frequency axis of the melody contour 
to 128 bins before feeding to the next layers.

\begin{figure*}[t]
\hspace{2.8cm}(a) 05-Winter.mp3 \hspace{2.3cm}(b) 07-Calypso.mp3 \hspace{2cm}(c) 01-Black Friday.mp3
\vspace{-0.4cm}
\begin{center}
\includegraphics[trim=0.2cm 3.3cm 1cm 1.9cm,clip,scale=0.45]{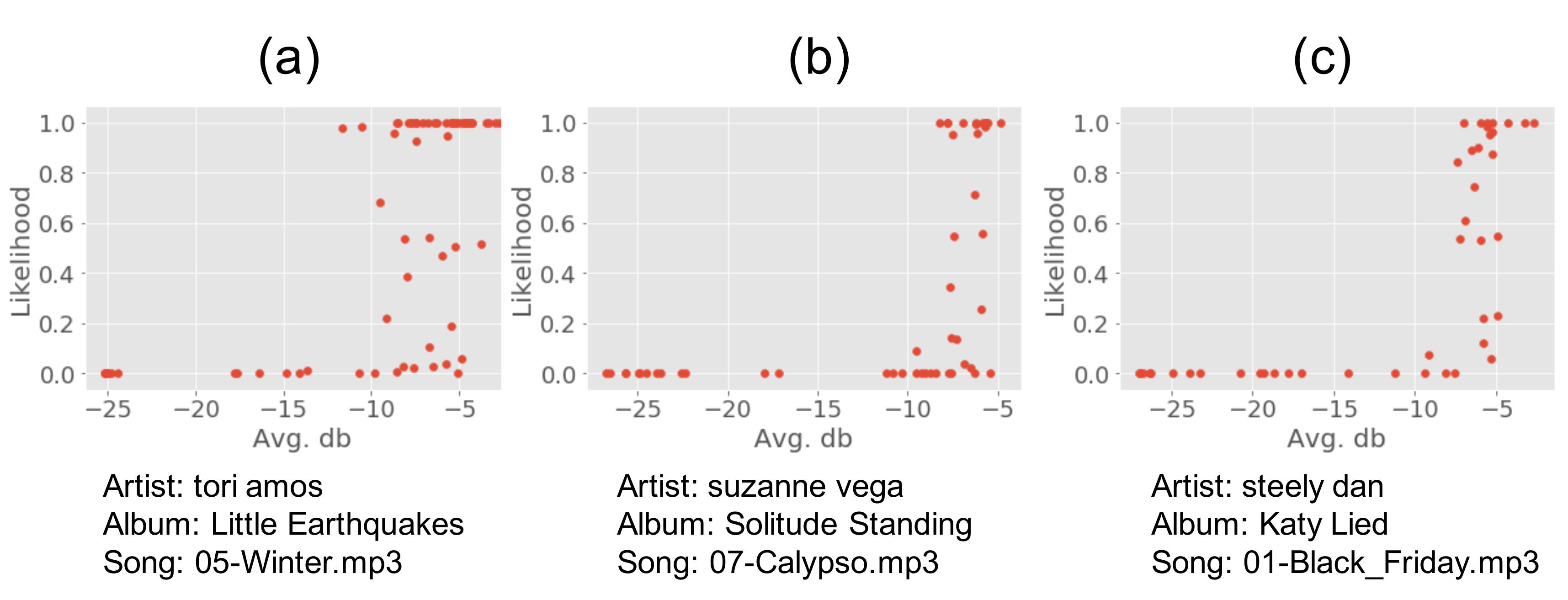}
\caption{The scatter plots showing the likelihood score for the correct singer of a testing song for different 5-sec segments of that song, predicted by the CRNN model trained on `Data aug.' The segments are sorted from left to right in each plot according to vocalness, the average decibel value of the vocal-separated part of that song. 
The three plots show the same trend: the model do predict the correct singer for the frames with average vocal db greater than $-10$, but not for the non-vocal frames. 
}
\label{fig:correlation}
\end{center}
\end{figure*}

\section{Experiments}
\label{sec:experiments}

The models are evaluated using artist20 \cite{artist20} under the album split, averaging the F1 scores of three independent runs. 




\begin{table}[t]\
\caption{Average testing F1 score on the artist20 dataset; note that `CRNN+Origin' resembles the model in \cite{nasrullah2019music}.}
\label{table:1}
\centering
\begin{tabular}{|c|l|c c|} 
 \hline
 Model  & Data &  F1~/~5-sec & F1~/~Song \\ [0.5ex] 
 \hline\hline
 \multirow{4}{4em}{CRNN} 
 & Origin    & 0.50 & 0.67 \\
 & Vocal-only    & 0.39 & 0.61\\
 & Remix     & 0.39 & 0.65 \\
 & Data aug. & 0.47 & 0.74 \\
 \hline
  \multirow{4}{4em}{CRNN$^{\dagger}$}
 & Origin    & \textbf{0.54} & 0.67 \\
 & Vocal-only    & 0.48 & 0.71 \\
 & Remix     & 0.46 & 0.71 \\
 & Data aug. & 0.50  & 0.74 \\
 \hline
 \multirow{4}{4em}{CRNNM} 
 & Origin    & 0.53 & 0.69 \\
 & Vocal-only    & 0.42 & 0.66\\
 & Remix     & 0.39 & 0.65\\
 & Data aug. & 0.45 & \textbf{0.75}\\
 \hline
 \end{tabular}
\end{table}

\subsection{Experimental results}
\label{ssec:experimental_result}

From Table~\ref{table:1}, we see that CRNNM performs the best among the three models.
This result shows  that using melody contour as additional features helps SID.

Our `CRNNM+Data aug' model achieves 0.75 song-level F1 score, which is greatly higher than that (0.67) obtained by the best existing model (`CRNN+Origin') \cite{nasrullah2019music} for artist20.

Table~\ref{table:1} also shows that, for all the three models,  training on \emph{Data aug} outperforms those trained on \emph{Origin} for the song-level result, validating the effectiveness of the data augmentation method.  We also note that, using Vocal-only performs even worse than using Origin for the case of CRNN and CRNNM, suggesting that the models trained with Origin may benefit from the additional (unwanted confounding) information in the accompaniment. Using Remix alone addresses this issue, but its result is no better than using Origin alone.  The combination of the three data (i.e., Origin, Vocal-only, and Remix) significantly boosts the song-level F1 score.

The F1 score at the 5-sec level is much worse than that at the song level, highlighting the importance of majority voting in aggregating the result. 
One important reason for this is the presence of non-vocal parts in a song. 
To demonstrate this, we regard ``vocalness'' as the mean volume of the vocal-separated clip for each 5-sec segment, and then compute the correlation between the vocalness and the prediction of the ground truth singer for test songs by our CRNN model trained with Data aug training set. The resulting correlation coefficient (0.39) indicates a weak relationship between these two factors.
Figure~\ref{fig:correlation} shows the result for three random test songs. We see that the model assigns high likelihood scores to the correct singer for the vocal frames (i.e., frames with larger avg db) but not for the non-vocal frames. 
We therefore suggest that 1) song-level accuracy is more important than 5-sec level accuracy, 2) future work may consider employing a vocal/non-vocal detector (e.g., \cite{dataaug-is-work1}) in both the training and testing stages.

\begin{figure}
\begin{center}
The t-SNE of CRNN \\
\end{center}
~~~~~~~~\includegraphics[trim=0.3cm 0.2cm 2cm 0.1cm,clip,scale=0.26]{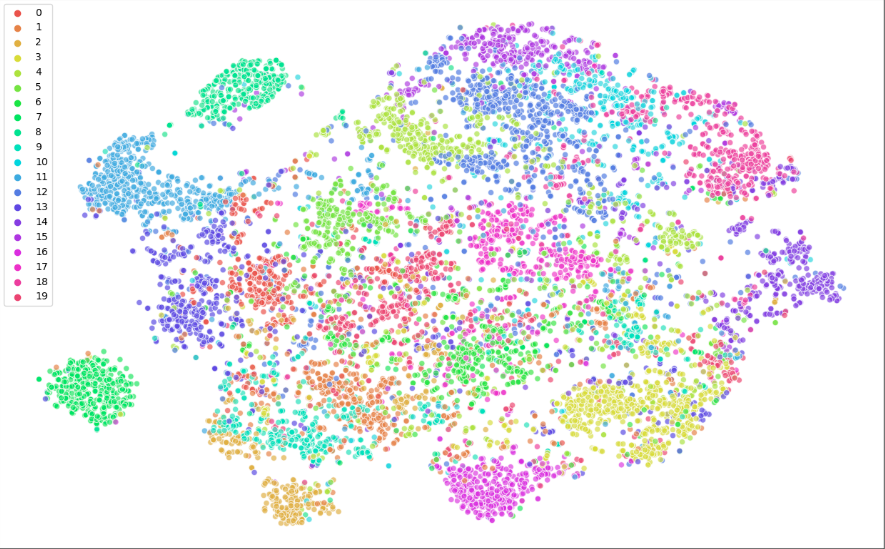} \\
\begin{center}
The t-SNE of CRNNM \\
\end{center}
\includegraphics[trim=2.5cm 4.0cm 0.1cm 1.5cm,clip,scale=0.26]{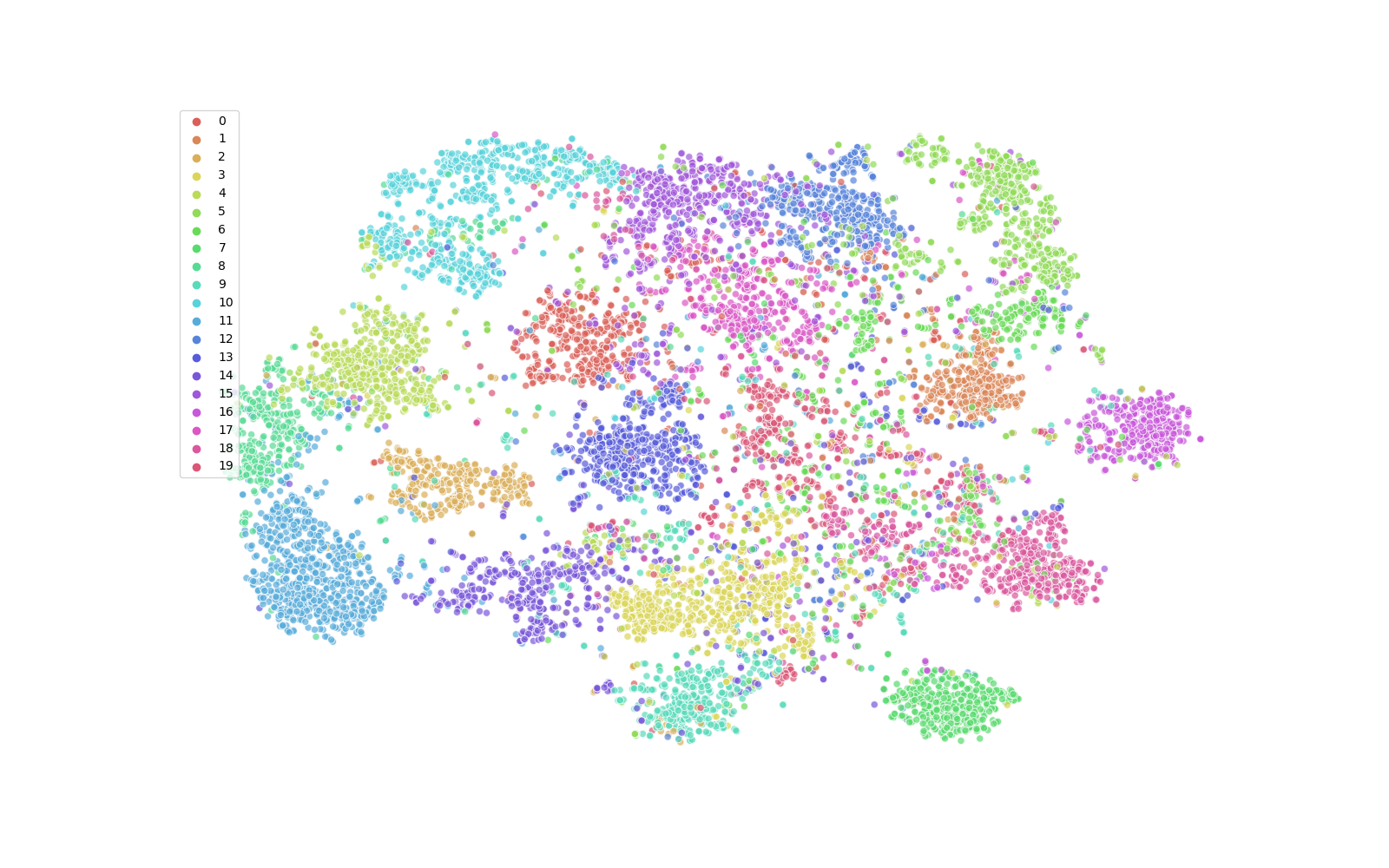}
\caption{Visualization of the embeddings (projected into 2D by t-SNE) generated by the models trained on the \emph{Origin} training set for the testing samples (5s segment; under the album split). Upper: the result of CRNN (i.e., the model shown in Figure \ref{fig:crnnm} but without the melody branch); lower: the result of CRNNM (i.e., the model shown in Figure \ref{fig:crnnm}). }
\label{fig:t-sne}
\end{figure}

\subsection{Visualization}
After training, 
we can regard the output of the final fully-connected layer as an embedding of the input data. Visualizing the representations can give us some ideas of the behaviour and performance of our SID models. Therefore, we employ t-Distributed Stochastic Neighbor Embedding (t-SNE)~\cite{t-sne} to project the computed embedding vectors to a 2-D space for visualization,
and to explore the structure of the predictions. For space limit, only the result of CRNN and CRNNM models trained on \emph{Origin} are presented. The audio samples of testing set are drew and colored according to the ground truth artist labels in Figure~\ref{fig:t-sne}.
It can be seen from the result of CRNNM that samples from different singers are fairly well-separated in the embedding space.\footnote{We note that similar visualization of the learned embedding space is also provided in \cite{nasrullah2019music}. Yet, they consider the song-split setting in their visualization, while we consider the more challenging yet realistic case of album-split. Therefore, although the embeddings shown in their work seem to be even more separated, we still consider the result here promising.}  The result of CRNN looks less separated, suggesting again that a model taking additional melody feature may do SID better.

\section{Conclusions}
\label{sec:conclusions}

The paper proposes a new SID model extending from CRNN and involving the use of melody information by leveraging CREPE \cite{kim2018crepe}. Also, a data augmentation method called shuffle-and-remix is adopted to avoid the confounds from the accompaniments by using source separation \cite{openunmix}. Our evaluation shows that both melody information and data augmentation improve the result, especially the latter. 
Future work includes three directions. First, to use a vocal detector \cite{dataaug-is-work1} as a pre-filter for SID. Second, to investigate replacing convolutions by GRUs for the melody branch since the melody contour is a time series. Lastly, to try other data augmentation methods such as pitch shifting, time stretching, or a shuffle-and-remix variant that considers the key and tempo  while remixing.







\bibliographystyle{IEEEbib}
\small
\bibliography{refs}

\end{document}